\documentstyle[12pt,epsf]{article}

\newcommand{\vspu}{\vspace*{6mm}}

\newcommand\be{\begin{equation}}
\newcommand\ee{\end{equation}}
\newcommand\bea{\begin{eqnarray}}
\newcommand\eea{\end{eqnarray}}

\makeatletter \@addtoreset{equation}{section} \makeatother

\begin{document}
\begin{center}
{\large\bf The $B_{s}\rightarrow K$ Form Factor in The Whole Kinematically Accessible
Range}
\vspu \\
\ \\
Zuo-Hong Li$^{\rm a,b,c,d,}$\footnote{Email: rxj@ytu.edu.cn}, Fang-Ying
Liang\protect\(^b\protect\), Xiang-Yao Wu\protect\( ^{c}\protect \) and Tao
Huang\protect\( ^{a,c}\protect \)
\\
\vspu \ \\
\footnotesize{a. CCAST (World Laboratory), P.O.Box 8730,Beijing 100080, China\\
\vspu
b. Department of Physics, Yantai University, Yantai 264005,
China\footnote{Mailing address}\\
\vspu
c. Institute of High Energy Physics, P.O.Box
918(4), Beijing 100039, China\\
\vspu
d. Department of Physics, Peking University,
Beijing 100871, China
\\
}
\end{center}
\vspu
\begin{center}{\large \bf Abstract}
\end{center}
\normalsize
%%%%%%%%%%%%%%%%%%%%%%%%%%%%%%%%%%%%%%%%%%%%%%
%%%%%%%%%%%%%%%%%%%%%%%%%%%%%%%%%%%%%%%%%%%%%%

A systematic analysis is presented of the $B_{s}\rightarrow K $ form factor $f\left(
q^{2}\right) $ in the whole range of momentum
transfer $q^{2}$, which would be useful to analyzing the future data on $%
B_{s}\rightarrow K$ decays and extracting $\left| V_{ub}\right| $. With a
modified QCD light cone sum rule (LCSR) approach, in which the contributions
cancel out from the twist 3 wavefunctions of $K$ meson, we investigate in
detail the behavior of $f\left( q^{2}\right) $ at small and intermediate $%
q^{2}$ and the nonperturbative quantity $f_{B^{\ast }}g_{B^{\ast }B_{s}K}$ $%
(f_{B^{\ast }}$ is the decay constant of $B^{\ast }$ meson and $g_{B^{\ast
}B_{s}K}$ the $B^{\ast }B_{s}K$ strong coupling), whose numerical result is
used to study $q^{2}$ dependence of $f\left( q^{2}\right) $ at large $q^{2}$
in the single pole approximation. Based on these findings, a pole model from
the best fit is formulated, which applies to the calculation on $f\left(
q^{2}\right) $ in the whole kinematically accessible range. Also, a
comparison is made with the standard LCSR predictions.

\begin{verse}
\end{verse}

\newpage
\baselineskip 20pt

A study on heavy-to-light exclusive processes plays a complementary role in
the determination of fundamental parameters of the standard model(SM) and in
the development of QCD theory. At present, an important task in the SM is to
extract the $CKM$ parameter $\mid {V_{ub}}\mid $. Recently, a new QCD
factorization formula \cite{s1} has been proposed for nonleptonic $B$
decays and has been applied to discuss phenomenology of $B\rightarrow \pi
\pi ,\pi K$ and $\pi D$. This approach, however, is not adequate to
extracting $\mid {V_{ub}}\mid $ from the relevant data, for the effects of
the long distance QCD are anyway difficult to control in exclusive
nonleptonic processes. Semileptonic B decays into a light meson, induced by $%
b\rightarrow u$ transition, are regarded as the most promising processes
suitable for such a purpose. Nevertheless, in this case precision extraction
of $\mid {V_{ub}}\mid $ requires a rigorous estimate of the relevant
hadronic matrix elements. It is a great challenge, because of our inability
to deal with nonperturbative QCD effects from the first principle. Heavy
quark symmetry is less predictive, and also lattice QCD calculation is
restricted to a certain kinematical region, for heavy-to-light decays.
Although QCD sum rule method has been being an effective QCD-based approach
to nonperturbative dynamics, with its extensive uses in phenomenology, the
resulting form factors behave very badly in the heavy quark limit $%
m_{Q}\rightarrow \infty $, the reason being that one omits the effects of
the finite correlation length between the quarks in the physical vacuum. In
order to overcome the disadvantage, an alternative to the ''classical'' sum
rule method, QCD light-cone sum rules (LCSR)\cite{s2}, has been developed
and has proven to be an advanced tool to deal with heavy-to-light
transitions. There exist a lot of its applications in the literature. For a
detailed description of this method, see \cite{s3}. However, a problem to
need solving is how to control the pollution by higher twist, especially
twist 3 wavefunctions, which are known very poorly and whose influence on
the sum rules is considerable in most cases. In \cite{s4,s5}, an improved
LCSR approach has been worked out, to eliminate twist 3 wavefunctions and
enhance the reliability of sum rule calculations, and has been applied to
reexamine heavy-to-light form factors in the region of momentum transfer $%
0\leq q^{2}\leq m_{b}^{2}-2m_{b}\Lambda _{QCD}$, where the operator product
expansion (OPE) goes effectively in powers of small light-cone distance $%
x^{2}$.

Most of previous works \cite{s3,s4,s6,s7,s8} are devoted to discussing $%
B\rightarrow \pi ,\rho $ semileptonic transitions within the context of
LCSR, with the aim to extract $\mid {V_{ub}}\mid $. A study on $%
B_{s}\rightarrow K$ semileptonic processes is equally important. As compared
with the case $B\rightarrow \pi ,\rho $, however, the $B_{s}\rightarrow K$
form factors are more difficult to evaluate, for SU(3) breaking corrections
to the twist 3 wave functions of $K$ meson have not been investigated
completely in the literature. Explicitly, this problem can be avoided in our
approach \cite{s4,s5}. On the other hand, to calculate the semileptonic
widths one must find another way to estimate the form factors at the large
momentum transfer $m_{b}^{2}-2m_{b}\Lambda _{QCD}\leq q^{2}\leq
(m_{B_{s}}-m_{K})^{2}$. In the letter, we investigate the $B_{s}\rightarrow K
$ form factor $f(q^{2})$ at the total momentum transfer with the improved
LCSR and a pole model.

Let us start with the following definition of the $B_{s}\rightarrow K$ form
factors $f(q^{2})$ and $\widetilde{f}(q^{2})$:
\begin{equation}
\langle K(p)|\overline{u}\gamma _{\mu }b|B(p+q)\rangle =2f(q^{2})p_{\mu }+%
\widetilde{f}(q^{2})q_{\mu }.
\end{equation}%
For $B_{s}\rightarrow K\ell \widetilde{\nu _{\ell }}$ transitions, as $%
l=e,\mu $ we can neglect the contributions from $\widetilde{f}(q^{2})$ due
to the smallness of $m_{e,\mu }$ and therefore only the form factor $%
f(q^{2}) $ is relevant. It can precisely be represented as

\begin{eqnarray}
f(q^{2}) &=&\frac{f_{B^{\ast }}g_{B^{\ast }B_{s}K}}{2m_{B^{\ast
}}(1-q^{2}/m_{B^{\ast }}^{2})}+\int\limits_{\sigma _{0}}^{\infty }\frac{\rho
(\sigma )d\sigma }{1-q^{2}/\sigma }  \nonumber \\
&=&F_{G}(q^{2})+F_{H}(q^{2}),
\end{eqnarray}%
with $f_{B^{\ast }}$ being the decay constant of $B^{\ast }$ meson, $%
m_{B^{\ast }}$ the $B^{\ast }$ meson mass, $g_{B^{\ast }B_{s}K}$ the strong
coupling defined by%
\begin{equation}
\langle B^{\ast }(q,e)K(p)|B_{s}(p+q)\rangle =-g_{B^{\ast }B_{s}K}(p\cdot e),
\end{equation}%
and $\rho (\sigma )$ a spectral function with the threshold $\sigma _{0}$.
Obviously, $F_{G}(q^{2})$ stands for the contribution from the ground state $%
B^{\ast }$ meson, which describes the principal behavior of $f(q^{2})$
around $q^{2}=q_{max}^{2}$, and $F_{H}(q^{2})$ parametrizes the higher state
effects in the $B^{\ast }$ channel. As we have known, the form factor $%
f(q^{2})$ may be estimated for the small and intermediate momentum transfers
by means of LCSR, and also the nonperturbative parameter $f_{B^{\ast
}}g_{B^{\ast }B_{s}K}$ is accessible within the same framework. Accordingly,
modelling the higher state contributions by a certain assumption and then
fitting (2) to its LCSR result $f_{LC}(q^{2})$ in the region accessible to
the light cone OPE, we might derive the form factor $f(q^{2})$ in the total
kinematical range to a better accuracy. For this purpose, we follow the
procedure in \cite{s4,s5} and consider a chiral current correlator used for
a LCSR sum rule calculation on $f_{LC}(q^{2})$ and $f_{B^{\ast }}g_{B^{\ast
}B_{s}K}$,
\begin{eqnarray}
\Pi _{\mu }(p,q) &=&i\int d^{4}xe^{iqx}\langle K(p)|T\{\overline{u}(x)\gamma
_{\mu }(1+\gamma _{5})b(x),\overline{b}(0)i(1+\gamma _{5})s(0)\}|0\rangle
\nonumber \\
&=&F(q^{2},(p+q)^{2})p_{\mu }+\widetilde{F}(q^{2},(p+q)^{2})q_{\mu }
\end{eqnarray}%
Inserting complete sets of the relevant intermediate states $|B^{H}\rangle $
in (4) and using the definition $\langle 0|\bar{s}i\gamma
_{5}b|B_{s}\rangle$\ $=\frac{m_{B_{s}}^{2}}{m_{b}+m_{s}}f_{B_{s}}$ and
$\langle
 0|\bar{u}\gamma _{\mu }b|B^{\ast }\rangle=m_{B^{\ast}}f_{B^{\ast
}}e_{\mu }$, we have the two hadronic representations of the invariant
function $F(q^{2},(p+q)^{2})$,

\begin{equation}
F_1^H(q^2,(p+q)^2)= \frac{2 f_{LC}(q^2) m_{B_s}^2 f_{B_s}}{(m_b+m_s)
(m_{B_s}^2-(p+q)^2)} +\int \limits_{s_0}^{\infty}\frac {\rho_1^H(s)}{%
s-(p+q)^2}ds
\end{equation}
\begin{eqnarray}
F_{2}^{H}(q^{2},(p+q)^{2}) &=&\frac{m_{B_{s}}^{2}m_{B^{\ast
}}f_{B_{s}}f_{B^{\ast }}g_{B^{\ast }B_{s}K}}{(m_{b}+m_{s})(m_{B^{\ast
}}^{2}-q^{2})(m_{B_{s}}^{2}-(p+q)^{2})}  \nonumber \\
&&+\int \int \frac{\rho _{2}^{H}(s_{1},s_{2})\Theta (s_{1}-s_{0}^{^{\prime
}})\Theta (s_{2}-s_{0})}{(s_{1}-q^{2})(s_{2}-(p+q)^{2})}{ds_{1}}{ds_{2}}.
\end{eqnarray}%
Several definite interpretations for (5) and (6) are in order. The two
dispersion integrals include, in addition to the contributions of the
resonances carrying the same quantum numbers as the corresponding ground
states in the pole terms, the effects due to the relevant orbit-excited $B$
mesons. Taking it into account that these orbit-excited states are far from
the lowest $B_{s}$ and $B^{\ast }$ mesons, and the lowest two of them are
slightly below the first excited $B_{s}$ and $B^{\ast }$ mesons in mass,
their contributions can effectively absorbed into a dispersion integral so
that thresholds $s_{0}$ and $s_{0}^{\prime }$ should correspond to the
squared masses of the lowest $0^{+}$ $B_{s}$ and $1^{+}$ $B$ mesons
respectively. On the other hand, the vector current $\overline{u}\gamma
_{\mu }b$ and axial-vector current $\overline{u}\gamma _{\mu }\gamma _{5}b$
couple also to $0^{+}$ and $0^{-}$ $B$ mesons, respectively, which should be
considered in (6). The invariant function, however, does not receive such a
contribution as we have checked. Therefore, it is safe to separate the
hadronic expression $F_{2}^{H}(q^{2},(p+q)^{2})$ into a pole term and a
dispersion integral.

The task left is to calculate the correlator in QCD theory in order to
obtain the desired sum rules. To this end, we work in the large space-like
momentum regions: $(p+q)^{2}\ll 0$ for the $f_{LC}(q^{2})$ case and $%
q^{2}\ll 0$, $(p+q)^{2}\ll 0$ for the $f_{B_{s}}f_{B^{\ast }}g_{B^{\ast
}B_{s}K}$ case, so that the light cone OPE can be used for the correlator
under consideration. After contracting the $b$ quark operators, we encounter
some nonlocal matrix elements, which can systematically be expanded in
powers of the deviation from the light cone $x^{2}=0$ by defining the
relevant light cone wavefunctions of $K$ meson classified in terms of twist.
The explicit parametrizations of all those can be found in the literature,
and here are no more given for simplicity. A long but straightforward
calculation yields the light-cone QCD form of the invariant function $%
F(q^{2},(p+q)^{2})$,

\begin{eqnarray}
F^{QCD}(q^{2},(p+q)^{2})&=&2f_{K}m_{b}\left\{ \int\limits_{0}^{1}{du}\left[
\frac{\varphi _{K}(u)}{m_{b}^{2}-(q+up)^{2}}-\frac{%
8m_{b}^{2}[g_{1}(u)-G_{2}(u)]}{[m_{b}^{2}-(q+up)^{2}]^{3}}+\frac{2ug_{2}(u)}{%
\left[ m_{b}^{2}-\left( q+up\right) ^{2}\right] ^{2}}\right] \right.
\nonumber \\
&&\left.+\int\limits_{0}^{1}d{\alpha }\int D\alpha _{i}\frac{2\varphi
_{\perp }(\alpha _{i})+2\tilde{\varphi}_{\perp }(\alpha _{i})-\varphi
_{\parallel }(\alpha _{i})-\tilde{\varphi}_{\parallel }(\alpha _{i})}{%
[m_{b}^{2}-(q+\beta p)^{2}]^{2}}\right\} ,
\end{eqnarray}%
to twist 4 accuracy. Here $\varphi _{K}(u)$ is the twist 2 wavefunction,
while the others have twist 4; the parameter $\beta =\alpha _{1}+\alpha
\alpha _{3}$ and $D\alpha _{i}=d\alpha _{1}d\alpha _{2}d\alpha _{3}\delta
\left( 1-\alpha _{1}-\alpha _{2}-\alpha _{3}\right).$ At this
point, we put once again an emphasis on that differently from the existing
LCSR calculations, the twist 3 wavefunctions make precisely a vanishing
contribution to the correlator we choose. This is essentially important to
enhancing precision of the LCSR calculation.

As usual, we need to make the Borel improvements on the theoretical
expression: $F^{QCD}(q^{2},(p+q)^{2})\rightarrow \bar{F_1}^
{QCD}(q^{2},M^{2})$, $F^{QCD}(q^{2},(p+q)^{2})\rightarrow \bar{F_2}^
{QCD}(M_{1}^{2},M_{2}^{2})$, and then match them onto the individual Borel
improved hadronic forms. \ Invoking the quark-hadron duality ansatz, the
final sum rules for $f_{LC}(q^{2})$ and $g_{B^{\ast }B_{s}K}$ read
respectively:
\begin{eqnarray}
f(q^{2})&=&\frac{m_{b}(m_{b}+m_{s})}{m_{B_{s}}^{2}f_{B_{s}}}f_{K}e^{\frac{%
m_{B_{s}}^{2}}{M^{2}}}\left\{ \int_{\Delta }^{1}\frac{du}{u}e^{-\frac{%
m_{b}^{2}-(q^{2}-um_{K}^{2})(1-u)}{uM^{2}}}\left[ \varphi _{K}\left(
u\right) \right. \right.  \nonumber \\
&&\left.-\frac{4m_{b}^{2}}{u^{2}M^{4}}g_{1}(u)+\frac{2}{uM^{2}}%
\int\limits_{0}^{u}g_{2}(v)dv\left( 1+\frac{m_{b}^{2}+q^{2}}{uM^{2}}\right) %
\right]  \nonumber \\
&&+\int\limits_{0}^{1}d\alpha \int D\alpha _{i}\frac{\Theta (\beta -\Delta )%
}{\beta ^{2}M^{2}}e^{-\frac{m_{b}^{2}-(q^{2}-\beta m_{K}^{2})(1-\beta )}{%
\beta M^{2}}}\left[ 2\varphi _{\perp }(\alpha _{i})+2\tilde{\varphi}_{\perp
}(\alpha _{i})-\varphi _{\parallel }(\alpha _{i})-\tilde{\varphi}_{\parallel
}(\alpha _{i})\right]  \nonumber \\
&&-4m_{b}^{2}e^{\frac{-s_{0}}{M^{2}}}\left[ \frac{1}{(m_{b}^{2}-q^{2})^{2}}%
\left( 1+\frac{s_{0}-q^{2}}{M^{2}}\right) g_{1}(\Delta )-\frac{1}{%
(s_{0}-q^{2})(m_{b}^{2}-q^{2})}\frac{dg_{1}(\Delta )}{du}\right]  \nonumber
\\
&&\left.-2e^{\frac{-s_{0}}{M^{2}}}\left[ \frac{m_{b}^{2}+q^{2}}{%
(s_{0}-q^{2})(m_{b}^{2}-q^{2})}g_{2}(\Delta )-\frac{1}{(m_{b}^{2}-q^{2})}%
\left( 1+\frac{m_{b}^{2}+q^{2}}{m_{b}^{2}-q^{2}}\left( 1+\frac{s_{0}-q^{2}}{%
M^{2}}\right) \right) \int\limits_{0}^{\Delta }g_{2}(v)dv\right] \right\} ,
\nonumber \\
&&
\end{eqnarray}

\begin{eqnarray}
f_{B_{s}}f_{B^{\ast }}g_{B^{\ast }B_{s}K} &=&\frac{2m_{b}(m_{b}+m_{s})f_{K}}{%
m_{B_{s}}^{2}m_{B^{\ast }}}e^{\frac{m_{B_{s}}^{2}+m_{B^{\ast }}^{2}}{2%
\overline{M}^{2}}}\left\{ \overline{M}^{2}\left[ e^{-\frac{m_{b}^{2}+\frac{1%
}{4}m_{K}^{2}}{\overline{M}^{2}}}-e^{-\frac{S_{0}}{\overline{M}^{2}}}\right]
\varphi _{K}(1/2)\right.  \nonumber \\
&&\left. +e^{-\frac{{m_{b}^{2}+\frac{1}{4}m_{K}^{2}}}{{\overline{M}^{2}}}}%
\left[ g_{2}(1/2)-\frac{4m_{b}^{2}}{\overline{M}^{2}}\left[
g_{1}(1/2)-\int\limits_{0}^{1/2}g_{2}(v)dv\right] \right. \right.  \nonumber
\\
&&\left. \left. +\int\limits_{0}^{1/2}d\alpha _{1}\int\limits_{1/2-\alpha
_{1}}^{1-\alpha _{1}}\frac{d\alpha _{3}}{\alpha _{3}}\left[ 2\varphi _{\perp
}(\alpha _{i})+2\widetilde{\varphi }_{\perp }(\alpha _{i})-\varphi
_{\parallel }(\alpha _{i})-\widetilde{\varphi }_{\parallel }(\alpha _{i})%
\right] \right] \right\} ,
\end{eqnarray}%
where $\Delta =\frac{m_{b}^{2}-q^{2}}{s_{0}-q^{2}-m_{K}^{2}}$ and $\overline{%
M}^{2}=\frac{M_{1}^{2}M_{2}^{2}}{M_{1}^{2}+M_{2}^{2}}$. In the derivation of
(9), we have taken $M_{1}^{2}=M_{2}^{2}$ due to the fact that $B_{s}$ and $%
B^{\ast }$ mesons are nearly degenerate in mass, which renders the continuum
subtraction reduce to a simple replacement $e^{-\frac{m_{b}^{2}+\frac{1}{4}%
m_{K}^{2}}{\bar{M}^{2}}}\rightarrow e^{-\frac{m_{b}^{2}+\frac{1}{4}m_{K}^{2}%
}{\bar{M}^{2}}}-e^{-\frac{S_{0}}{\bar{M}^{2}}}$ for the leading twist 2 term.

Turning now to the numerical discussions on the sum rules. $B$ channel
parameters entering the sum rules are the $b$ quark mass $m_{b}$, $B$ meson
masses $m_{B_{s}}$ and $m_{B^{\ast }}$, decay constants $f_{B_{s}}$ and $%
f_{B^{\ast }}$, and threshold parameter $s_{0}$. We take $m_{B_{s}}=5.369\
GeV$, $m_{B^{\ast }}=5.325\ GeV$ and $m_{b}=4.8\ GeV$. As for the decay
constants $f_{B_{s}}$ and $f_{B^{\ast }}$, we have to reanalyze them in the
two-point QCD sum rule approaches \cite{s4,s5} with chiral current
correlators, to keep a consistency with the sum rules in question. The
results are found to be $f_{B_{s}}=0.142\ GeV$ and $f_{B^{\ast }}=0.132\ GeV$%
, as the threshold parameter $s_{0}=34\ GeV^2$ corresponding to the mean value
of squared masses of the lowest $0^{+}$ $B_{s}$ and $1^{+}$ $B$ mesons. For
the decay constant of $K$ meson and mass of $s$ quark, we use $f_{K}=0.16\
GeV$ and $m_{s}=0.15\ GeV$. The important point is to specify the set of the
light-cone wave functions of $K$ meson. Unlike the case of $\pi $ meson,
SU(3) breaking effects need considering for the distribution amplitudes of $K
$ meson. In the work, we use the model presented in \cite{s9} for the
leading twist wavefunction, which is based on an expansion over orthogonal
Gegenbauer polynomials with coefficients determined by means of QCD sum
rules. The explicit expression is

\begin{equation}
\varphi _{K}(u)=6u(1-u)\left\{ 1+1.8\left[ (2u-1)^{2}-\frac{1}{5}\right]
-0.5(2u-1)\left[ 1+1.2[(2u-1)^{2}-\frac{3}{7}]\right] \right\} ,
\end{equation}%
at the scale $\mu _{b}=\sqrt{m_{B_{s}}^{2}-m_{b}^{2}}$, measuring the mean
virtuality of the $b$ quark. For the twist-4 wave functions, we neglect the
SU(3) breaking effects and utilize the same forms as those of $\pi $ meson
investigated in \cite{s10}.

Having fixed the input parameters, one must look for a reliable range of the
Borel parameters $M^{2}$ and $\overline M^{2}$, which can be determined by
the standard procedure. The fiducial intervals are found to be $8\
GeV^{2}\leq M^{2}\leq 17\ GeV^{2}$, depending slightly on $q^{2}$, for $%
q^{2}=0-17\ GeV^{2}$ and $5\ GeV^{2}\leq \overline{M}^{2}\leq 10\ GeV^{2}.$
In the two "windows", the twist 4 wavefunctions contribute less than $9\%$
and $7\%$, and the continuum states at the levels lower than $25\%$ and $%
22\% $, respectively. The sum rule results for $f_{LC}(q^{2})$ show a weak
dependence on $M^{2}$ up to $q^{2}=17\ GeV^{2}$, varying between $\pm
3\%-\pm 5\%$ relative to their central values. For the product $%
f_{B_{s}}f_{B^{\ast }}g_{{B^{\ast }B_{s}K}}$, the resulting sum rule is $%
f_{B_{s}}f_{B^{\ast }}g_{{B^{\ast }B_{s}K}}=0.55\ GeV^{2}$, the uncertainty
due to $\overline{M}^{2}$ being $\pm 4\%$. Taking its central value, we get $%
g_{{B^{\ast }B_{s}K}}=29$. To evaluate better the $B$ pole contribution in
(2), however, we would give a direct sum rule result for $f_{B^{\ast }}g_{{%
B^{\ast }B_{s}K}}$, which can be obtained utilizing the analytic form
instead of the numerical result for the two-point sum rule for $f_{B_{s}}$
in (9). The result is $f_{B^{\ast }}g_{{B^{\ast }B_{s}K}}=3.57-4.19\ GeV$,
depending on the Borel parameters. The sum rule prediction $f_{LC}(q^{2})$,
together with that from the $B^{\ast }$ pole approximation, is illustrated
in Fig. 1. It is explicitly demonstrated that a perfect match between them
appears at $q^{2}\approx 15-20\ GeV^{2}$.

The influence on the sum rules should be investigated in detail from several
important sources of uncertainty: the twist 2 distribution amplitude $%
\varphi _{K}(u)$, $b$ quark mass $m_{b}$, decay constants $f_{B_{s}}$ and $%
f_{B^{\ast }}$, and threshold parameter $s_{0}$. Concerning the light cone
wavefunction $\varphi _{K}(u)$, there are some determinations other than
that in (10) in the literature. To investigate the sensitivity of the sum
rules to the choice of the non-asymptotic coefficients in $\varphi _{K}(u)$,
we consider the two models suggested in \cite{s7} and \cite{s8}, and
confront the resulting sum rules with our ones. If adopting $\varphi _{K}(u)$
in \cite{s7}, the resulting changes amount to $-8\%--9\%$ for the $%
f_{LC}(q^{2})$ case and to $\pm 5\%$ for the $f_{B^{\ast }}g_{B^{\ast
}B_{s}K}$ case. The almost same situation exists for that used in \cite{s8}.
 Therefore, the uncertainties caused by $\varphi _{K}(u)$ may be estimated
at a considerably small level. As for the $B$ channel parameters $m_{b}$, $%
f_{B_{s}}$, $f_{B^{\ast }}$ and $s_{0}$, considering a correlated variation
in the individually allowed ranges would give sufficient information on the
uncertainties induced by them. This can be done in such a way where letting $%
m_{b}$ vary from $4.7$ to $4.9\ GeV$, we observe the behavior of $%
f_{LC}(q^{2})$ and $f_{B^{\ast }}g_{B^{\ast }B_{s}K}$ by requiring that the
relevant decay constants take only the best fitting values. We find that
such an effect amounts to $6\%$ and $5\%$, respectively. At present, the
total uncertainties in $f_{LC}(q^{2})$ and $f_{B^{\ast }}g_{B^{\ast }B_{s}K}$
can respectively be estimated to be $20\%$ and $18\%$, by adding linearly up
all the considered errors.

It is important and interesting to make a comparison of our sum rule results
and those from the standard LCSR based on the corrector of vector and
pseudoscalar currents, which are easy to obtain using the twist-3
wavefunctions suggested in \citation{s11}, leaving the twist-4 distribution
amplitudes unchanged and making a corresponding replacement of the other
relevant input parameters in (79) and (44) of the second reference in \cite
{s6}. We observe that the standard approach gives the same matching range
as in our case and the resulting deviations from our predictions turn out to
be between $-10\%--15\%$, depending on $q^{2}$, in the total kinematically
accessible region. This denotes that both approaches are essentially
compatible with each other within the available errors.

With the yielded findings we would give a specific parametrization for $%
f(q^2)$ applicable to the whole kinematical region, which is helpful for the
future practical application. Assuming the higher state contribution in (2)
to obey $F_H(q^2)=a/\left( 1-bq^2/m_{B^{*}}^2-cq^4/m_{B^{*}}^4\right) $, we
have a pole model for $f(q^2)$,
\begin{equation}
f(q^{2})=\frac{f_{B^{\ast }}g_{B^{\ast }B_{s}K}}{2m_{B^{\ast
}}(1-q^{2}/m_{B^{\ast }}^{2})}+\frac{a}{1-bq^{2}/m_{B^{\ast
}}^{2}-cq^{4}/m_{B^{\ast }}^{4}}.
\end{equation}%
The parameter $a$ can easily be fixed at $-0.07$, using the central values
of $f_{LC}(0)$ and $f_{B^{\ast }}g_{B^{\ast }B_{s}K}$. In the region $%
q^{2}=0-18\ GeV^{2}$, the best fit of (11) to $f_{LC}(q^{2})$ yields $b=1.11$%
, and $c=-8.33$. The resulting $q^{2}$ dependence of $f(q^{2})$ is
demonstrated in Fig.1 too, for a comparison. It turns out that the fitting
results reproduce precisely the LCSR prediction up to $q^{2}=18\ GeV^{2}$
and support considerably the single pole description of the $%
B_{s}\rightarrow K$ form factor $f(q^{2})$ at large $q^{2}$.

Also, it is worthwhile to look roughly into SU(3) breaking effects in
heavy-to-light decays by considering the ratio of the derived $%
B_{s}\rightarrow K$ form factor over the corresponding $B\rightarrow \pi $
one. The $B\rightarrow \pi $ form factor has already been obtained for small
and intermediate $q^{2}$ in the improved LCSR approach in \cite{s4}.
Using all the same method as in present case, we can understand its behavior
at large $q^{2}$ and further get a parametrization holding for the total
kinematical range. For the common kinematical region to the two processes,
the resulting ratios, a comparable result $1.05-1.15$ with that from the
standard approach, favor a small SU(3) breaking effect.

We have given a detailed discussion on the $B_{s}\rightarrow K$ form factor $%
f(q^{2})$ in the whole kinematical region. To avoid the contamination with
the twist 3 wavefunctions, in which SU(3) breaking corrections have not been
analyzed systematically, an improved LCSR approach with some kind of chiral
current correlator has been applied to estimate the form factor $%
f_{LC}(q^{2})$ at small and intermediate $q^{2}$. The nonperturbative
quantity $f_{B^{\ast }}g_{B^{\ast }B_{s}K}$, an important input in the $%
B^{\ast }$ pole model for $f(q^{2})$, has also been calculated within the
same framework and the sum rule result has been adopted to study the
behavior of $f(q^{2})$ at large $q^{2}$. We find that the resulting $%
f_{LC}(q^{2})$ matches quite well with the estimate from the $B^{\ast }$
pole model at $q^{2}=15-20\ GeV^{2}.$ A comparison shows that our
predictions are in basic agreement with those from the standard LCSR. Based
on our findings, a pole model for $f(q^{2})$ has been worked out, which is
applicable to the total kinematically accessible region. The results
presented here would be used as analyzing the future data on $%
B_{s}\rightarrow K$ decays and extracting $\mid {V_{ub}}\mid $. A future
lattice calculation of the $B_s\rightarrow K$ form factors, which is available
for large $q^{2}$, will provide a direct test of our predictions. The same
approach applies also to discuss other heavy-to-light processes.

This work is in part supported by the National Science Foundation of China.

\newpage
\begin{figure}
\centerline{ \epsfxsize=12cm \epsfbox{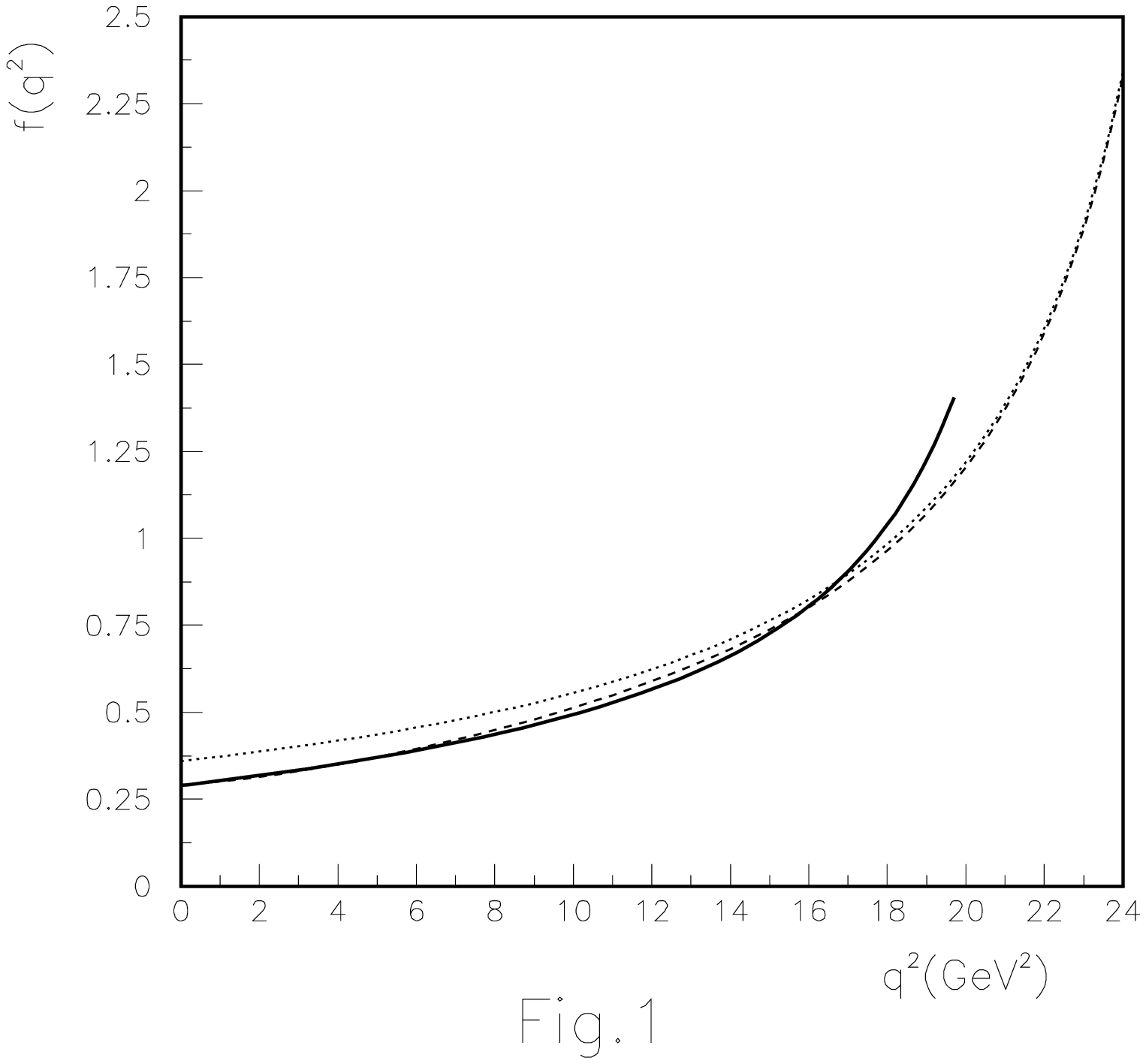} } \caption{\baselineskip 20pt The
$B_s\rightarrow K$ form factor $f(q^2)$ in the total kinematical range. The solid line
denotes the LCSR result $f_{LC}(q^2)$, which is reliable for $0\leq q^2 \leq 17\
GeV^2$. The dotted line expresses the $B^*$ pole prediction suitable for large $q^2$.
The best fit of Eq. (11) to $f_{LC}(q^2)$ is illustrated by the dashed line. It should
be understood that the plotted curves correspond to the central values of all the
relevant parameters.}
\end{figure}

\end{document}